\documentclass[]{aa} 

\usepackage{hyperref}
\addto\extrasenglish{
    
}
\addto\extrasenglish{
    
}
\hypersetup{
    colorlinks = true,
    linkcolor = {purple},
    citecolor = {blue}
}
\usepackage{graphicx}
\usepackage{txfonts}
\newcommand{\Msun}{M$_{\odot}$}

\newcommand{\lppr}{\stackrel{<}{\scriptstyle \sim}}
\newcommand{\lappr}{\raisebox{-0.4ex}{$\lppr$}}

\newcommand{\mesa}{MESA}

\newcommand{\cv}{CV}
\newcommand{\cvs}{CVs}

\newcommand{\msun}{M$_\odot$}

\defcitealias{rappaportetal83-1}{RVJ}

\begin{document} 

\definecolor{purple}{rgb}{0.4, 0.0, 0.8}

   \title{Further evidence of saturated, boosted, and disrupted magnetic braking from evolutionary tracks of cataclysmic variables}

\titlerunning{Saturated, boosted, and disrupted magnetic breaking for CVs}

   \author{Joaquín A. Barraza-Jorquera
          \inst{1}\textsuperscript{$\star$}
          \and
          Matthias R. Schreiber 
          \inst{1}
          \and
          Diogo Belloni
          \inst{1,2}
          }

\authorrunning{Barraza-Jorquera et al.}

   \institute{
   Departamento de F\'isica, Universidad T\'ecnica Federico Santa Mar\'ia, Avenida Espa\~na 1680, Valpara\'iso, Chile
              \and
              São Paulo State University (UNESP), School of Engineering and Sciences, Guaratinguetá, Brazil
              \\ \\ \textsuperscript{$\star$}\email{joaquin.barrazaj@usm.cl}
             }

   \date{Received XX, XXXX; accepted XX, XXXX}

  \abstract
   {
   Angular momentum loss through magnetic braking drives the spin-down of low-mass stars and the orbital evolution of a variety of close binary stars. Current theories for magnetic braking, often calibrated for one particular type of system, predict angular momentum loss rates that differ by several orders of magnitude. A unified prescription, even if fully empirical, would provide important constraints on the relation between angular momentum loss, stellar dynamos, and stellar magnetic activity.}
   {Recent studies have shown that a saturated, boosted, and disrupted (SBD) magnetic braking prescription explains the increase in the fraction of close systems among white dwarf plus M-dwarf binaries at the fully convective boundary, the period distribution of main-sequence star binaries, and the mass distribution of close M-dwarf companions to hot subdwarfs. With the aim of analyzing whether this prescription is also applicable to related binaries, we investigated the evolution of cataclysmic variables assuming a SBD magnetic braking prescription.}
   {We incorporated the SBD magnetic braking model into the stellar evolution code MESA and simulated the evolution of cataclysmic variables, testing different values for the boosting ($K$) and the disruption ($\eta$) parameters for different stellar parameters.}
   {The model accurately reproduces the mass transfer rates and the donor star mass-radius relation. The corresponding evolutionary tracks are in good agreement with the observed boundaries of the orbital period gap as well as the period minimum when assuming $K\simeq\eta\simeq30-50$. These values for $K$ and $\eta$ are slightly smaller than but consistent with those determined from detached binaries ($K\simeq\eta\,\gtrsim\,50$). }
   {Angular momentum loss through SBD magnetic braking can explain not only observations of detached binaries but also cataclysmic variables, that is, it is the only prescription currently available that is suitable for several types of close binary stars. 
   The model needs to be tested further in the context of other close binary and single stars, and the currently used semi-empirical convective turnover time for main-sequence stars needs to be replaced with self-consistent turnover times.}

   \keywords{
             binaries: close --
             methods: numerical --
             stars: evolution --
             white dwarfs
               }

   \maketitle

\section{Introduction}

Magnetic fields of low-mass stars force the mass lost in winds to corotate with 
the star up to the Alfv{\'e}n radius. This causes the terminal specific angular momentum of the wind to be higher 
than the specific angular momentum at the stellar surface. 
The resulting angular momentum loss is called magnetic wind braking and represents a fundamental ingredient 
of stellar astrophysics. 
Magnetic braking drives the spin-down of single stars \citep{schatzman62-1,mestel68-1} 
and the secular evolution of virtually all close binary stars \citep[see][for a recent review]{belloni+schreiber23-1}. 
Unfortunately, a unified theory describing angular momentum loss through magnetic braking as a function of stellar parameters does not exist. Instead, current prescriptions 
are intrinsically empirical and have been calibrated to reproduce either the spin-down of single stars or the orbital evolution of 
one type of close binary star. The predicted angular momentum loss rates of the different prescriptions therefore differ by several orders of magnitude \citep{kniggeetal11-1,belloni+schreiber23-1} and the proposed dependences on stellar parameters are fundamentally different. 

Decades ago, \citet{skumanich72-1} found that the rotational periods of Sun-like stars are proportional to the square root 
of their age, which implies that angular momentum loss should increase with the rotational speed to the power of three. 
However, chromospheric activity, coronal X-ray emission, flare activity, and magnetic field strengths in low-mass main-sequence stars are all correlated and strongly increase with rotation only up to a mass-dependent critical rotation rate. For shorter periods, the relation between activity and rotation saturates 
\citep[e.g.,][]{maguaddaetal20-1,reinersetal09-1,medinaetal20-1}.
Assuming that these observables relate with magnetic braking implies a much shallower  dependence of the torque on the spin period below a given rotation period \citep[e.g.,][]{chaboyeretal95-1,sillsetal00-1,andronovetal03-1}. 

In close binary stars with orbital periods shorter than ${\sim10}$\,days, tidal forces cause the stellar rotation to be synchronized with the orbital motion \citep[e.g.,][]{flemingetal19-1}. A Skumanich-like magnetic braking prescription therefore predicts very strong orbital angular momentum loss in these close binaries 
\citep[][hereafter RVJ]{rappaportetal83-1}. 
These high angular momentum loss rates for close binaries represent a key ingredient for the standard evolution model of cataclysmic variables (\cvs). 
According to this scenario, \cvs~with donor stars that still contain a radiative core experience strong angular momentum loss due to magnetic braking, which causes the donor stars to be bloated and the mass transfer rates to be high. When the secondary star becomes fully convective, at an orbital period of ${\sim3}$\,hours, magnetic braking becomes much weaker, which allows the donor star to shrink and to detach from its Roche lobe. The binary evolves as a detached system toward shorter periods until the mass transfer rate resumes at a much lower rate at an orbital period of ${\sim 2}$\,hours. 

This scenario is able to explain key features of the observed population of \cvs, most importantly the observed dearth of systems in the orbital period range between $\sim2$ and $\sim3$ hours, the so-called period gap \citep[e.g.,][]{schreiberetal24-1}, and the radii of donor stars derived from observations \citep{kniggeetal11-1}.
Furthermore, observations of \cv~progenitors strongly support the idea of significantly reduced magnetic braking at the fully convective boundary. The fraction of close binaries that must have evolved through common envelope evolution among observed white dwarfs with M-dwarf companions strongly increases for fully convective companion stars, indicating that these systems take much more time to evolve into \cvs~after they emerge from the common envelope phase \citep{schreiberetal10-1}. In addition, evidence of detached \cvs~crossing the period gap has been provided \citep{zorotovicetal16-1}. 

Further and independent support comes from the fact that the period gap is much less pronounced for \cvs~containing strongly magnetic white dwarfs. This observational fact confirms that the origin of the period gap is related to magnetism \citep{bellonietal20-1,schreiberetal24-1}, most likely a reduction in magnetic braking when the white dwarf magnetic field is strong enough to connect to the magnetic field of the donor star. The coupling of the two magnetospheres reduces the wind zones of the donor star and in turn the rate at which angular momentum is carried away from the binary \citep[see also][]{webbink+wickramasinghe02-1}. This reduced magnetic braking model is also consistent with observations of detached close binaries comprising magnetic white dwarfs. In these objects, higher rates of wind accretion
onto the strongly magnetic white dwarfs is observed, which implies that magnetic white dwarfs can capture a considerably greater portion of the wind material from their companion stars \citep{parsonsetal21-1}.

For decades, the evolution of the rotation of single stars was usually described by relatively inefficient saturated magnetic braking prescriptions, while for close binaries much more efficient braking laws such as those proposed by RVJ were used.
The idea of using braking laws with very different dependences on the rotation rate for single stars and those in close binaries was 
recently shattered by the finding that observations of close main-sequence binary stars strongly support saturated magnetic braking prescriptions \citep{elbadryetal22-1}. As a possible way out of this dilemma, \citet{bellonietal24-1} suggested a magnetic braking prescription that predicts strong enough angular momentum loss for stars that still have a radiative core, a drastic decrease at the fully convective boundary, and a period dependence that includes saturation. 
Here we call this prescription saturated, boosted, and disrupted (SBD) magnetic braking. It has been shown that SBD magnetic braking can explain the observations presented by \citet{schreiberetal10-1} as well as the mass distribution of the companions in hot subdwarfs \citep{blombergetal24-1} and the period distribution of main-sequence binaries \citep{bellonietal24-1}. We tested whether this prescription produces evolutionary tracks for \cvs~that are in agreement with the observed period distribution and mass-radius relation of \cv~donor stars. 
\section{Calibrated saturated, boosted, and disrupted magnetic braking in \mesa}

We used the Modules for Experiments in Stellar Astrophysics (\mesa) code \citep[][version 24.03.1]{Paxton2011, Paxton2013, Paxton2015, Paxton2018, Paxton2019, Jermyn2023} to compute the evolution of \cvs~to test the prescription of magnetic braking suggested by \citet{bellonietal24-1}. 
{MESA solves the coupled time-dependent equations that govern the structure and composition of evolving stars and is therefore capable to calculate fully 
consistent evolutionary tracks of \cvs.}

The \mesa~equation of state is a blend of the OPAL \citep{Rogers2002}, SCVH
\citep{Saumon1995}, FreeEOS \citep{Irwin2004}, HELM \citep{Timmes2000},
PC \citep{Potekhin2010}, and Skye \citep{Jermyn2021} equations of state.
Radiative opacities are primarily from OPAL \citep{Iglesias1993,
Iglesias1996}, with low-temperature data from \citet{Ferguson2005}
and the high-temperature, Compton-scattering dominated regime by
\citet{poutanen17-1}. Electron conduction opacities are from
\citet{Cassisi2007} and \citet{blouinetal20-1}.
Nuclear reaction rates are from JINA REACLIB \citep{Cyburt2010}, NACRE \citep{Angulo1999}, and
additional tabulated weak reaction rates \citet{Fuller1985, Oda1994,
Langanke2000}. Screening is included via the prescription of \citet{Chugunov2007}.
Thermal neutrino loss rates are from \citet{Itoh1996}.

We otherwise used standard assumptions for \cv~evolution. 
Roche lobe radii in binary systems are computed using the fit of
\citet{Eggleton1983}. Mass transfer rates in Roche lobe
overflowing binary systems were determined following the
prescription of \citet{Ritter1988}.
Systemic angular momentum loss through the emission of gravitational was included as described in \citet[][]{Paczynski1967}. 
We approximated the white dwarf as a point mass and assumed its mass to remain constant, that is, that the same amount of mass that is accreted during a nova cycle is expelled during the eruption in rough agreement with model predictions \citep[e.g.,][]{yaronetal05-1}.

It has been shown by \citet{schreiberetal16-1} that consequential angular momentum loss (CAML) can cause \cvs~with low-mass white dwarfs to evolve into dynamically unstable mass transfer, which brings the predicted and observed white dwarf mass distributions of \cvs~into agreement and leads 
to a substantially reduced predicted space density, one that is much closer to values derived from observations \citep{bellonietal18-1}.
We therefore assume the empirical relation for CAML (hereafter eCAML) proposed by \citet{schreiberetal16-1}, that is, 
\begin{equation}
    \frac{\dot{J}_{\text{CAML}}}{J} = \nu \frac{\dot{M}_2}{M_2},
    \label{eq:eCAML}
\end{equation}
where $M_2$ is the donor mass, $\dot{M_2}$ the mass transfer rate, and $\nu$ depends only on the white dwarf mass:
\begin{equation}
    \nu(M_1) = \frac{C}{M_1},
    \label{eq:nu_eCAML}
\end{equation}
with $C = 0.35$ \msun. 
To test the impact of the adopted prescription for CAML, we also ran some simulations assuming that the material expelled during a nova eruption carries the specific angular momentum of the white dwarf. In this case, often called classical CAML, $\nu$ depends on both stellar masses:
\begin{equation}
    \nu = \frac{M_2^2}{M_1(M_1 + M_2)}.
    \label{eq:nu_classicalCAML}
\end{equation}

The saturated and disrupted magnetic braking prescription calibrated by \citet{bellonietal24-1} is based on the first 
saturated magnetic braking law proposed by \citet{chaboyeretal95-1}: 

\begin{equation}
\dot{J}_{\rm SAT} \, = \,
-\beta
\left(
  \frac{R_2}{{\rm R}_{\odot}} \,
  \frac{{\rm M}_{\odot}}{M_2}
\right)^{1/2}
\left\{
\begin{array}{ll}
\Omega_2^3,                     & {\rm if} \; \Omega_2 \leq \Omega_{\rm crit}, \\
\Omega_2\,\Omega_{\rm crit}^2,  & {\rm if} \; \Omega_2 > \Omega_{\rm crit},
\end{array}
\right.
\label{eq:MBsat}
\end{equation}
\noindent
where ${\beta=2.7\times10^{47}}$~erg~s$^{-1}$ \citep{andronovetal03-1}, and $M_2$, $R_2$, and $\Omega_2$ are the mass, radius, and spin frequency (in s$^{-1}$) of the main-sequence star, respectively.
$\Omega_{\rm crit}$ is the critical angular velocity above which saturation occurs and is assumed to be 

\begin{equation}
\Omega_{\rm crit} \ = \ 
10\,\Omega_\odot
\left(
  \frac{\tau_{\odot}}{\tau_{\rm 2}}\,
\right),
\label{eq:SCRIT}
\end{equation}

\noindent
where ${\Omega_\odot=3\times10^{-6}~{\rm s}^{-1}}$ and $\tau_{\rm 2}$ is the empirical convective turnover timescale of main-sequence stars
from \citet{wrightetal11}, that is, 

\begin{equation}
\log_{10}
\left(
 \frac{\tau_{\rm 2}}{\rm d}
\right)
=
1.16
-1.49
\log_{10}
\left( 
  \frac{M_2}{{\rm M}_\odot}
\right)
-0.54
\log_{10}^2
\left( 
  \frac{M_2}{{\rm M}_\odot}
\right).
\label{eq:TAU}
\end{equation}

The lower the mass of the main-sequence star, the longer the convective turnover timescale and the longer the critical spin period below which magnetic braking is saturated.
The saturation spin period is ${\sim21.6}$~d for a $0.1$~\Msun~star and ${\sim2.85}$~d for a $0.9$~\Msun~star. 
This means that \cvs~and their progenitors are always in the saturated regime. 

To test different strengths of magnetic braking and different degrees of disruption at the fully convective boundary, \citet{bellonietal24-1} introduced two multiplicative factors, that is, a factor $K$ that scales the strength of magnetic braking and a factor $\eta$ by which magnetic braking is reduced for fully convective stars leading to a prescription given by  
\begin{equation}
\dot{J}_{\rm MB} \, = \,
\left\{
\begin{array}{ll}
K \, \dot{J}_{\rm SAT} \, ,         & {\rm (radiative~core + convective~envelope)}, \\
 & \\
\left(K \, \dot{J}_{\rm SAT}\right)/\eta \, , & {\rm (fully~convective}).
\end{array}
\right.
\label{eq:MBrecipe}
\end{equation}
\citet{bellonietal24-1} found that both the fraction of post common envelope binaries among all white dwarf plus M-dwarf binaries \citep{schreiberetal10-1} and the period distribution of close main-sequence binaries \citep{elbadryetal22-1} can be explained 
if both $\eta$ and $K$ exceed ${\sim50}$.

For comparison, we also calculated several evolutionary tracks assuming the standard \citetalias{rappaportetal83-1} magnetic braking prescription for \cvs:
\begin{equation}
\dot{J}_{\rm RVJ} \approx -3.8 \times 10^{-30} \, M \, R_{\odot}^4 \, \left(\frac{R}{R_{\odot}}\right)^{\gamma} \omega^3. 
\label{eq:RVJ}
\end{equation}
Here $M$ is the mass of star (in g) and $\omega$ is the rotational angular frequency (in s$^{-1}$) of the main-sequence star.

\section{Evolutionary tracks} 

The most important observational constraints on any theoretical model of \cv~evolution are the measured donor star radii \citep{kniggeetal11-1,mcallisteretal19-1} and the observed orbital period gap \citep{kniggeetal11-1,schreiberetal24-1}. 
Before we compare evolutionary tracks based on the SBD magnetic braking prescription with  
these observables for different values of the parameters $K$ and $\eta$, test the impact of eCAML
as well as different metallicities of the donor star and different white dwarf masses, we compare the predictions of SBD magnetic braking to evolutionary tracks calculated with \citetalias{rappaportetal83-1} standard disrupted magnetic braking 
using the normalization factors derived by \citet{kniggeetal11-1}.

 \begin{figure*}
    \centering
        \includegraphics[width=1.0
    \linewidth]{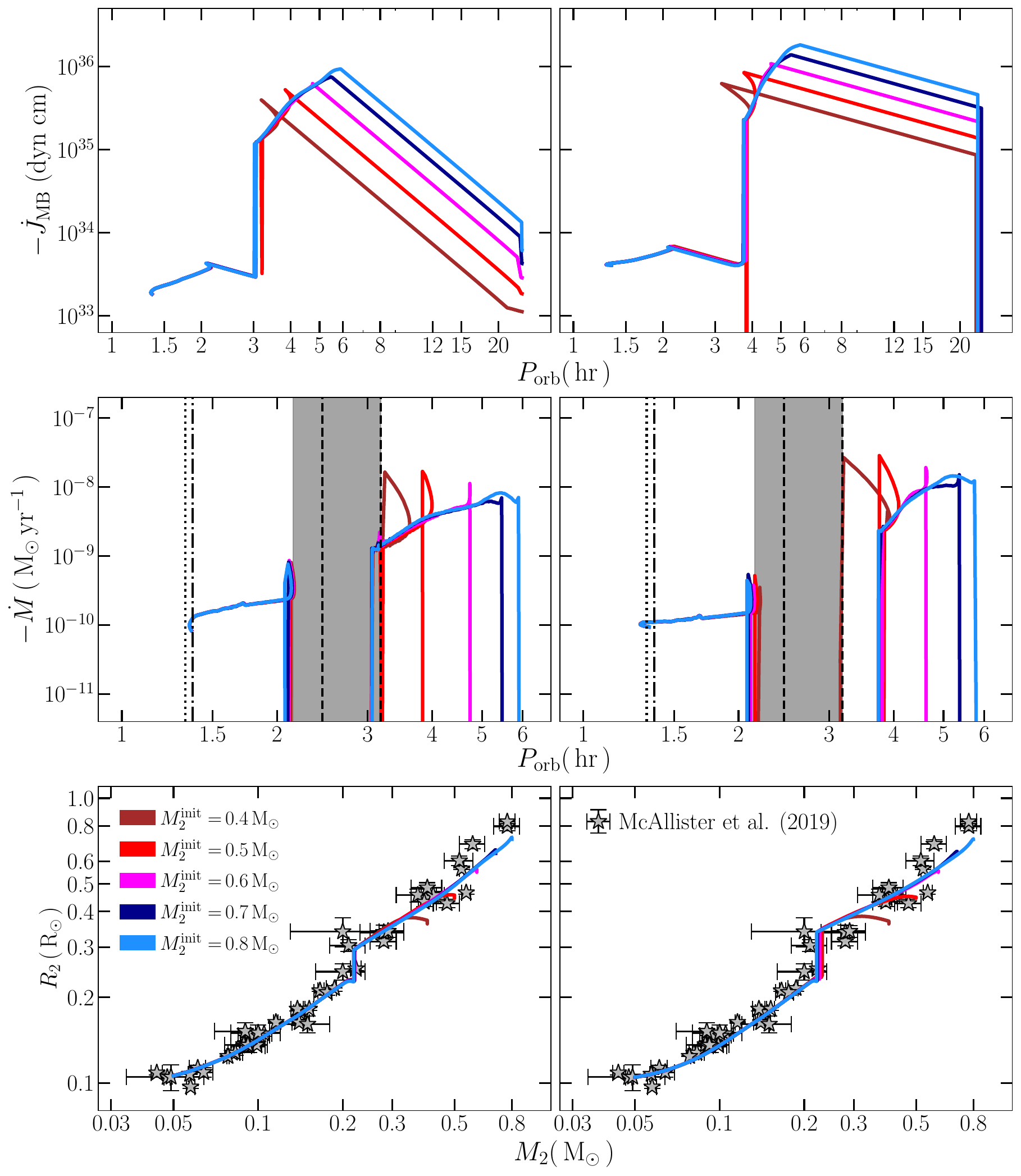} 
     \caption{The calibrated SBD magnetic braking prescription can reproduce the period gap for cataclysmic variables. Left panels: Evolution of \cvs\  using standard \citetalias{rappaportetal83-1} magnetic braking (Eq. \ref{eq:RVJ}) assuming $\gamma = 3$. Right panels: Using SBD magnetic braking (Eq. \ref{eq:MBrecipe}) assuming $K = \eta = 50$ as in \citet{bellonietal24-1}. Angular momentum loss due to magnetic braking (top), mass transfer rate (middle), and the mass-radius relation (bottom) for \cvs\  with different initial donor masses represented by the colors. The boundaries of the period gap derived by \citet[][]{schreiberetal24-1} are indicated by the dashed lines (at periods of $147 - 191 \, {\rm min}$) and those estimated by \citet[][]{kniggeetal11-1} as the gray shaded area (covering $129 - 191 \,{\rm min}$). The period minimum was estimated by \citet[][]{kniggeetal11-1} to be $P_{\rm min} \approx 82.4 \, {\rm min}$ (dotted-dashed line) and $79.6\pm0.2$\,min (dotted line) by \citet{mcallisteretal19-1}. In the bottom panels the markers represent the masses and radii of CVs measured by \citet{mcallisteretal19-1} and their respective error bars. The initial parameters of our \cv~models were fixed at $M_{\rm WD} = 0.83$ \msun~and $P_{\rm orb} = 1$\,d.}
     \label{fig:SBD+RVJ_Jdot+Mdot}
 \end{figure*}

\subsection{Comparing the predictions of RVJ and SBD models for magnetic braking}
\label{Sect:RVJvsSBD}

The top and middle panels of Fig. \ref{fig:SBD+RVJ_Jdot+Mdot} show the evolution of the angular momentum loss rates through magnetic braking and the mass transfer rate predicted for cataclysmic variable systems considering the standard \citetalias{rappaportetal83-1} magnetic braking model (left panels) with the normalization factors suggested by \citet{kniggeetal11-1} and the SBD magnetic braking prescription (right panels). In both cases we assumed eCAML \citep{schreiberetal16-1}.
In the bottom panels of Fig. \ref{fig:SBD+RVJ_Jdot+Mdot} we compare the predictions of both models with measurements of masses and radii of \cv~donors. Evolutionary tracks are shown for different initial donor masses and fixed values $K=\eta=50$. 

The first obvious difference, already noted by \citet{bellonietal24-1}, concerns the pre-CV phase, that is, the evolution from the initial post common envelope period to the onset of stable (angular momentum loss driven) mass transfer. The much stronger angular momentum loss rates predicted by the SBD model reduce the duration of this phase from $\approx\,8.5$\,Gyr to $\approx$\,112 Myr (depending on the assumed donor mass and for our fixed initial separation). This difference will have a strong impact on the predicted number of \cvs~with evolved donor stars and the total number of predicted \cvs. Quantifying this impact requires performing binary population synthesis, which is beyond the scope of this paper. Here, as a first step, we investigated the evolutionary paths of \cvs~assuming angular momentum loss through SBD magnetic braking.  

Independent of the used magnetic braking prescription, the donor star radius determines the orbital period at which mass transfer starts. For all donor masses we assumed, the mass transfer starts above the period gap. The onset of mass transfer for the lowest donor mass we assumed occurs just at the upper edge of the period gap (the brown track in Fig. \ref{fig:SBD+RVJ_Jdot+Mdot}). 

\begin{figure*}
    \centering
        \includegraphics[width=1.0
    \linewidth]{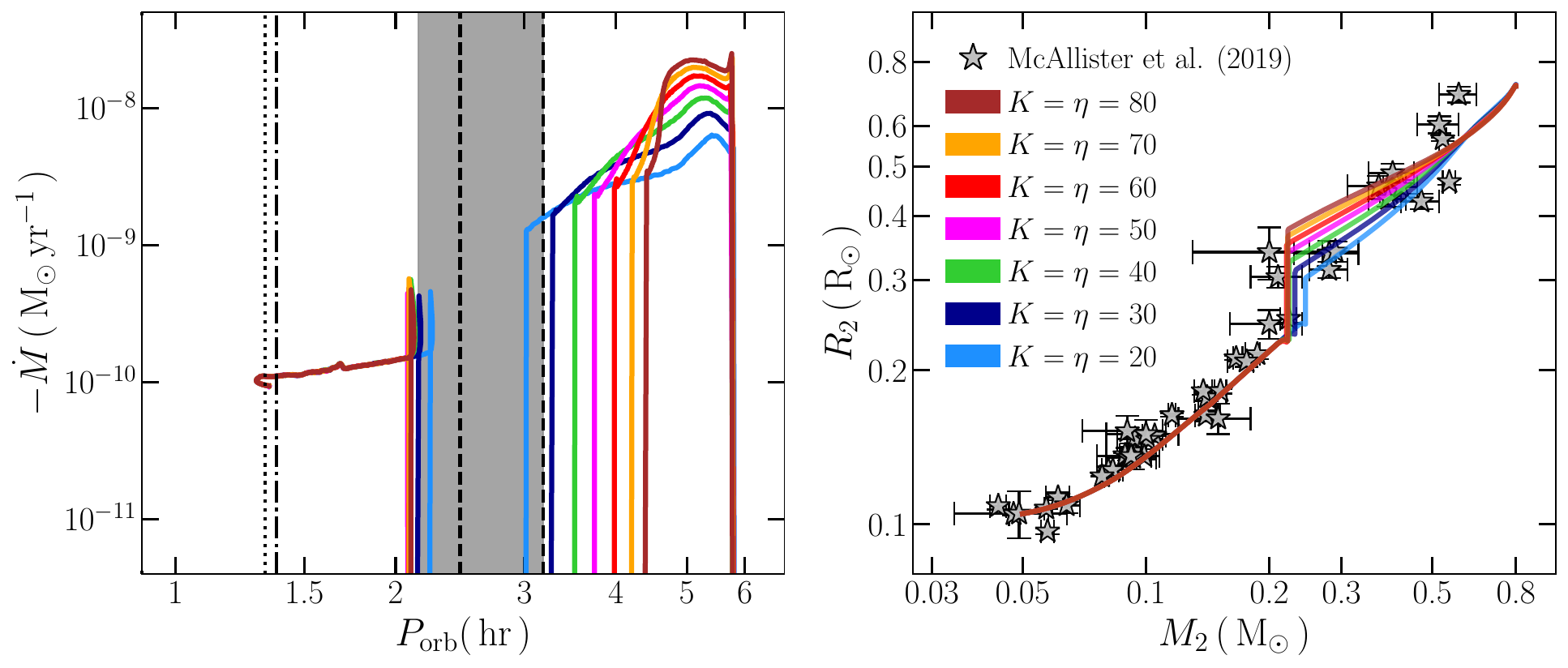} 
     \caption{Stronger SBD magnetic braking leads to a longer and earlier detached phase (period gap). The colors represent different values of the SBD magnetic braking parameters $K$ and $\eta$ assuming that $K = \eta$. We set the initial donor mass to $M_2 = 0.8$ \msun~. The black lines and gray area, the initial values for $M_{WD}$ and $P_{orb}$, and the symbols are the same as in Fig. \ref{fig:SBD+RVJ_Jdot+Mdot}.}
     \label{fig:Mdot+R2_K=eta}
 \end{figure*}

 \begin{figure*}
    \centering
        \includegraphics[width=1.0
    \linewidth]{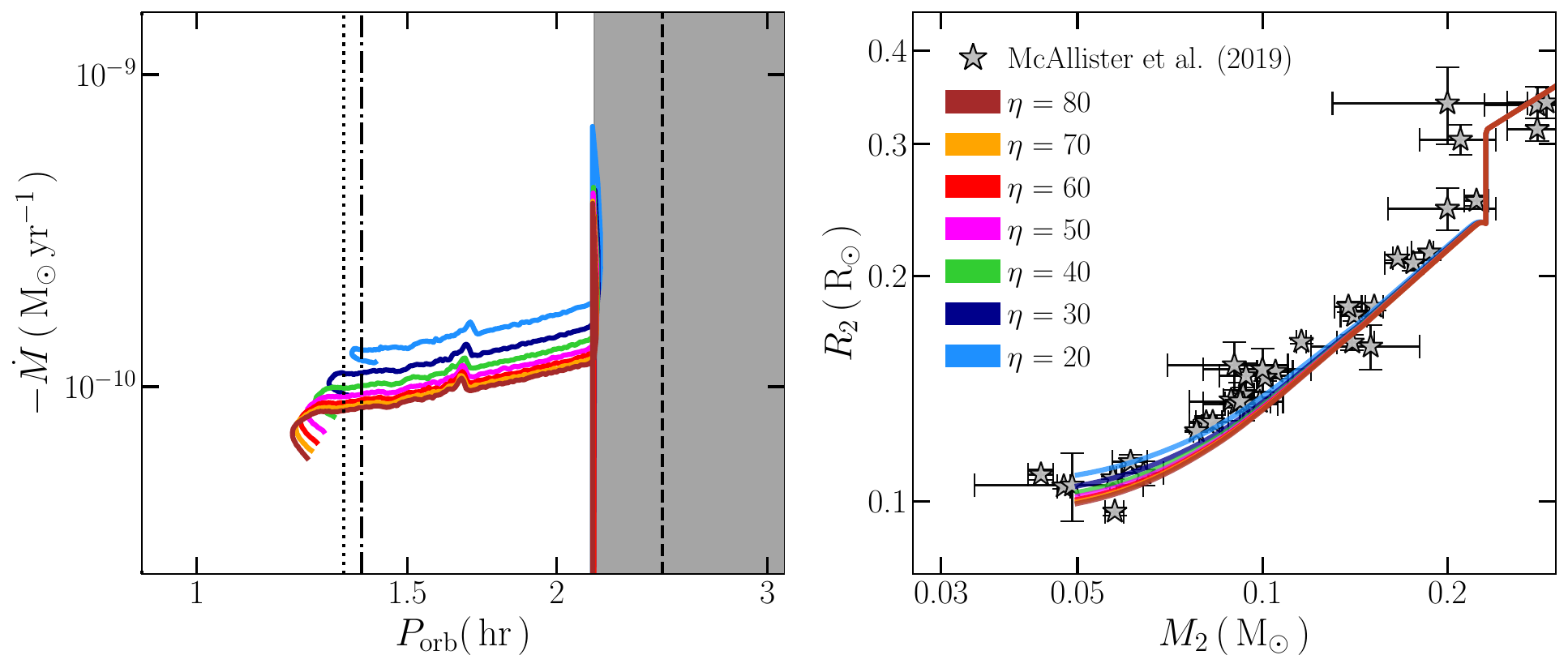}
     \caption{Similar to Fig. \ref{fig:Mdot+R2_K=eta} but with the boosting parameter fixed to $K=30$ and with different values of $\eta$ represented by the different colors. Both panels show the evolution of the system below the period gap. Lower values of $\eta$ (stronger SBD magnetic braking below the gap) produce slightly higher mass transfer rates and donor star radii in the fully convective case. This leads to an increase in the predicted period minimum.}
     \label{fig:Mdot+R2_K30_eta}
 \end{figure*}

As soon as the donor star becomes fully convective, both models predict a detached phase because the donor has time to shrink.
The angular momentum loss rates predicted by the SBD magnetic braking model above the gap are somewhat larger than those predicted by the standard \citetalias{rappaportetal83-1} prescription, which causes a larger increase in the donor star radius (Fig. \ref{fig:SBD+RVJ_Jdot+Mdot}, bottom panels). The shorter mass transfer timescale for stronger braking causes the donor star to become fully convective at a slightly lower mass. The combination of the two effects (lower mass but more bloated for stronger braking) leads binaries with stronger angular momentum loss rates (SBD model) to detach at a slightly longer orbital period (Fig. \ref{fig:SBD+RVJ_Jdot+Mdot}, middle panels). 
The small difference in donor mass at the beginning of the period gap implies that for both models mass transfer starts again more or less at the same orbital period (Fig. \ref{fig:SBD+RVJ_Jdot+Mdot}, middle panels). 

Both models reproduce reasonably well  the measured donor star mass-radius relation (Fig. \ref{fig:SBD+RVJ_Jdot+Mdot}, bottom right) for donor masses below $0.6$\,\Msun~but   
both magnetic braking models fail at matching the observed data at higher donor star masses, that is, $M_{2} \gtrsim 0.6$ \msun. 
Both models predict the lower edge of the orbital period gap to be very similar to the $P \approx 129\,{\rm min}$ measured by \citet[][]{kniggeetal11-1}, 
which is well below the value recently measured by \citet{schreiberetal24-1}.  
This probably indicates that the SDSS sample of \cvs~presented by \citet{inightetal23-1,inightetal23-2} is still biased against low mass transfer (and short orbital period) systems. Alternatively (or in addition), the orbital period range between $\sim2-2.5$\,hr might be populated by \cvs~that are born in the gap, which is expected to affect the lower boundary derived from observed samples but is not considered in our tracks.  

{In both models the secular evolution of \cvs~below the period gap is assumed to be driven by angular momentum loss due to residual magnetic braking and gravitational radiation. In the case of the \citetalias{rappaportetal83-1} model, we followed \citet{kniggeetal11-1} and assumed an angular momentum loss equal to $1.47$ times that of gravitational radiation due to residual magnetic braking.   
For the SBD model magnetic braking reduces to the 
original saturated magnetic braking prescription below the gap 
as long as $K=\eta$. }
In both cases the tracks provide reasonable fits to the observed orbital period minimum of $P_{\rm min} \approx 76-82 \, min$ \citep{kniggeetal11-1,mcallisteretal19-1}. 
{While for residual magnetic braking the fit to the observed period minimum is almost perfect -- which is no surprise as scaling the corresponding angular momentum loss to $1.47$ times that of gravitational radiation} was invented to fit the period minimum -- the SBD model seems to predict slightly too inefficient angular momentum loss as the predicted period minimum is slightly shorter than the observed one. 

The most significant difference concerns the onset of the detached phase when the donor star becomes fully convective. In the case of the \citetalias{rappaportetal83-1} model the fit to the observed edge of the period gap
(corresponding to 191 min)
is almost perfect.  
In contrast, the SBD model for $K=\eta=50$ predicts the beginning of the detached phase to occur at a longer orbital period ($P\approx224$\,min). 
However, so far we have tested the SBD model only for one white dwarf mass, solar metallicity and for ${K=\eta=50}$. In what follows we test how the predictions of the model depend on these parameters.

\subsection{Impact of the parameters $K$ and $\eta$}
\label{sec:k-eta}
 
In Fig. \ref{fig:Mdot+R2_K=eta} we show how the parameters $K$ and $\eta$ affect the mass transfer rate and the mass-radius relation of the donor star under the condition $K=\eta$. This condition implies that the evolution below the period gap is not affected because 
the two parameters cancel each other out 
(Eq.\,\ref{eq:MBrecipe}) and angular momentum loss therefore always corresponds to the standard saturated magnetic braking prescription. 

In contrast, as long as the donor star still has a radiative core, that is, before the period gap, the strength of magnetic braking is affected by $K$ and varying this parameter significantly changes the evolutionary tracks. 
Most importantly, the gap starts earlier if $K$ is larger. The observed upper edge of the period gap is well reproduced for $K=30$. 
The mass-radius relation is consistent with the values derived from observations in the range $M_{2} \lesssim 0.5$ \msun~and for $K\lesssim 50$. Independent of the assumed value of $K$, the three largest radii measured by \citet{mcallisteretal19-1} are slightly larger than predicted by the models.  
 
To test the effects of the parameter $\eta$ on the evolution of 
\cvs~with fully convective donor stars, that is, below the period gap, we fixed $K = 30,$ which provides the best fit for the upper edge of the period gap (Fig. \ref{fig:Mdot+R2_K=eta}), and varied $\eta$ from $20$ to $80$. 
Higher values of $\eta$ imply a stronger reduction of magnetic braking at the fully convective boundary.  
The strongest angular momentum loss through magnetic braking below the gap, and therefore the largest mass transfer rates, are obtained for $\eta = 20$. This value produces a period bounce at an orbital period similar to the observed period minimum (Fig. \ref{fig:Mdot+R2_K30_eta}). For larger values of $\eta$ the donor is less driven out of thermal equilibrium and therefore less bloated, which causes the period minimum to be predicted at orbital periods shorter than observed.

\subsection{Impact of the white dwarf mass and eCAML}

\begin{figure*}[h]
    \centering
        \includegraphics[width=1.0\linewidth]{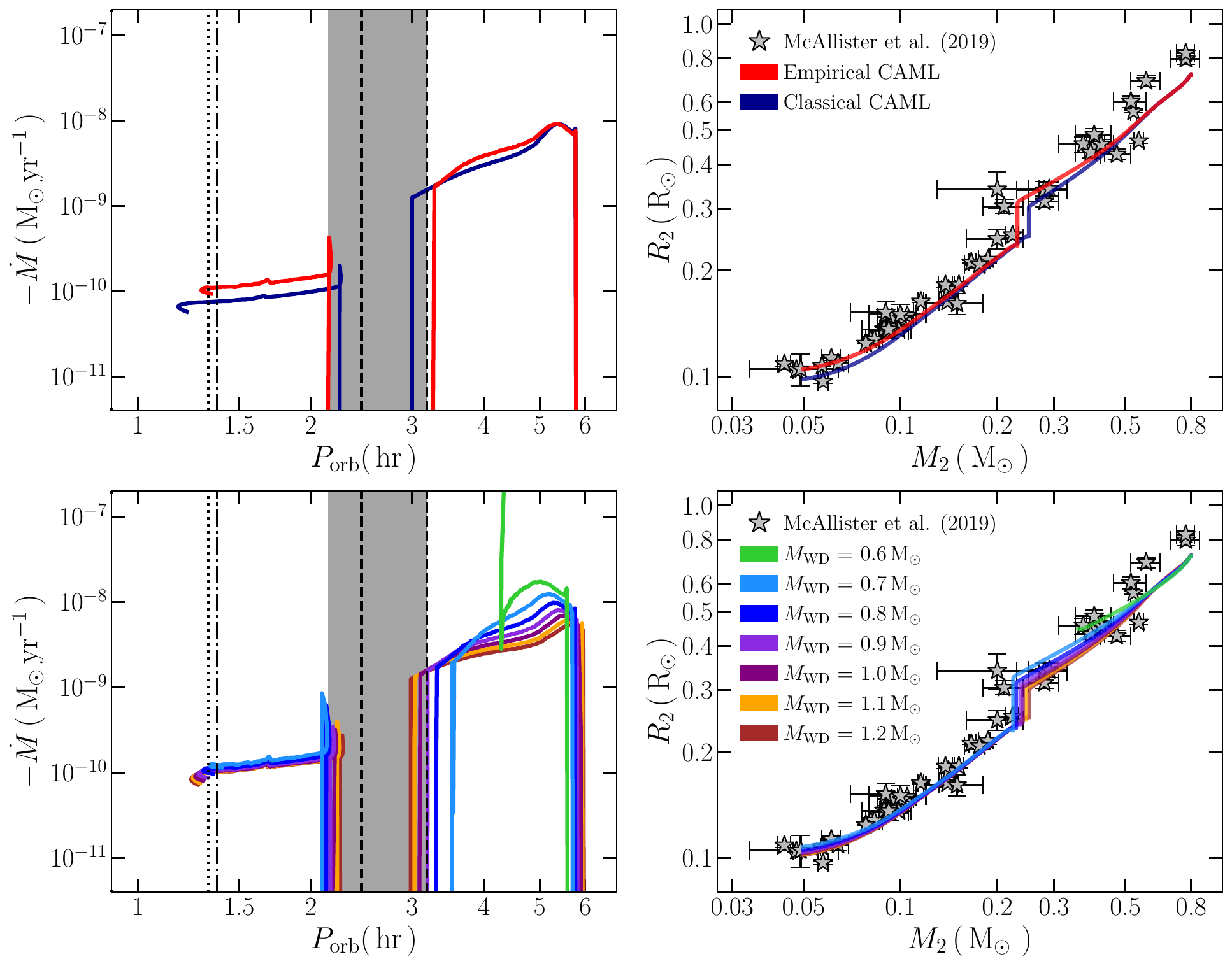}
     \caption{Predictions of the model for different white dwarf masses (bottom panels) and assumed consequential angular momentum (CAML) loss prescriptions (top panels). For the average white dwarf mass in \cvs,~incorporating eCAML (red) produces an earlier detachment (beginning of the period gap) than assuming classical CAML (top panels). The period range covered by the detached phase decreases for more massive white dwarfs (bottom-left panel). The parameters used to calculate the evolutionary tracks were $M_2 = 0.8$ \msun, $P_{\mathrm{orb}} = 1\,d,$ and $K=\eta=30$. Symbols are the same as in Fig. \ref{fig:SBD+RVJ_Jdot+Mdot}.}
     \label{fig:Mdot+R2_MWD}
\end{figure*}

In the previous sections we assumed a fixed white dwarf mass corresponding roughly to the average white dwarf mass ($M_{\rm WD}=0.83$ \msun) in \cvs~\citep{zorotovicetal11-1, palaetal22-1}. 
However, in reality, we expect a distribution of white dwarf masses for the population of \cvs. We therefore here evaluate the impact of the white dwarf mass on the evolutionary tracks. This is particularly important because the assumed prescription for angular momentum loss generated by mass transfer (eCAML; see Eqs. \ref{eq:eCAML} and \ref{eq:nu_eCAML}) postulates a strong dependence of this angular momentum loss on the white dwarf mass. In fact, differences in evolutionary tracks for different white dwarf masses are largely caused by the dependence of eCAML on the white dwarf mass.

The bottom panels of Fig. \ref{fig:Mdot+R2_MWD} show that for smaller white dwarf masses, the mass transfer rate increases until mass transfer becomes dynamically unstable for a white dwarf mass of $0.6$\,\Msun. 
The onset of the period gap varies as expected, that is, for stronger angular momentum loss (lower white dwarf mass), the detached phase starts at a longer period. However, for the lowest white dwarf mass we tested ($0.6$\,\Msun), mass transfer becomes dynamically unstable and the two stars merge before the system reaches the period gap. The prediction of these merger events for \cvs~with lower mass white dwarfs is the reason for the excellent agreement between predicted and observed white dwarf mass distributions if eCAML is incorporated \citep{schreiberetal16-1,palaetal22-1}. 

However, it is important to remember that eCAML represents an ad hoc prescription for angular momentum loss caused by mass transfer and that the physical mechanism behind this additional angular momentum loss is at least unclear. While it appears likely that the white dwarf mass distribution is indeed shaped by unstable mass transfer caused by CAML, largely because of independent evidence of this scenario \citep[][]{schreiberetal16-1,zorotovic+schreiber17-1,palaetal22-1}, whether eCAML correctly describes angular momentum loss caused by mass transfer for \cvs~that remain stable is less certain. For example, instead of the inverse linear dependence on white dwarf mass given by Eq. \ref{eq:nu_eCAML}, a steeper or a more moderate dependence that provide equally good fits to the white dwarf mass distribution might exist. 

Because of these uncertainties related to eCAML, we compare in the top panels of Fig. \ref{fig:Mdot+R2_MWD} the predictions of the SBD magnetic braking prescription with eCAML and with classical CAML. The latter assumes that the angular momentum carried away in nova eruptions corresponds to the specific angular momentum of the white dwarf \citep[see][for more details]{schreiberetal16-1}.  
Clearly, the choice of the CAML prescription may affect the values of $K$ and $\eta$ that best match the observed data. The larger mass transfer rates predicted by eCAML cause the donor star to become fully convective at a slightly longer period but lower mass, resulting in a longer period gap. This suggests that a greater value of $K$ and $\eta$ would be in better agreement with observations if classical CAML is assumed.

\begin{figure*}[h]
    \centering
        \includegraphics[width=1.0\linewidth]{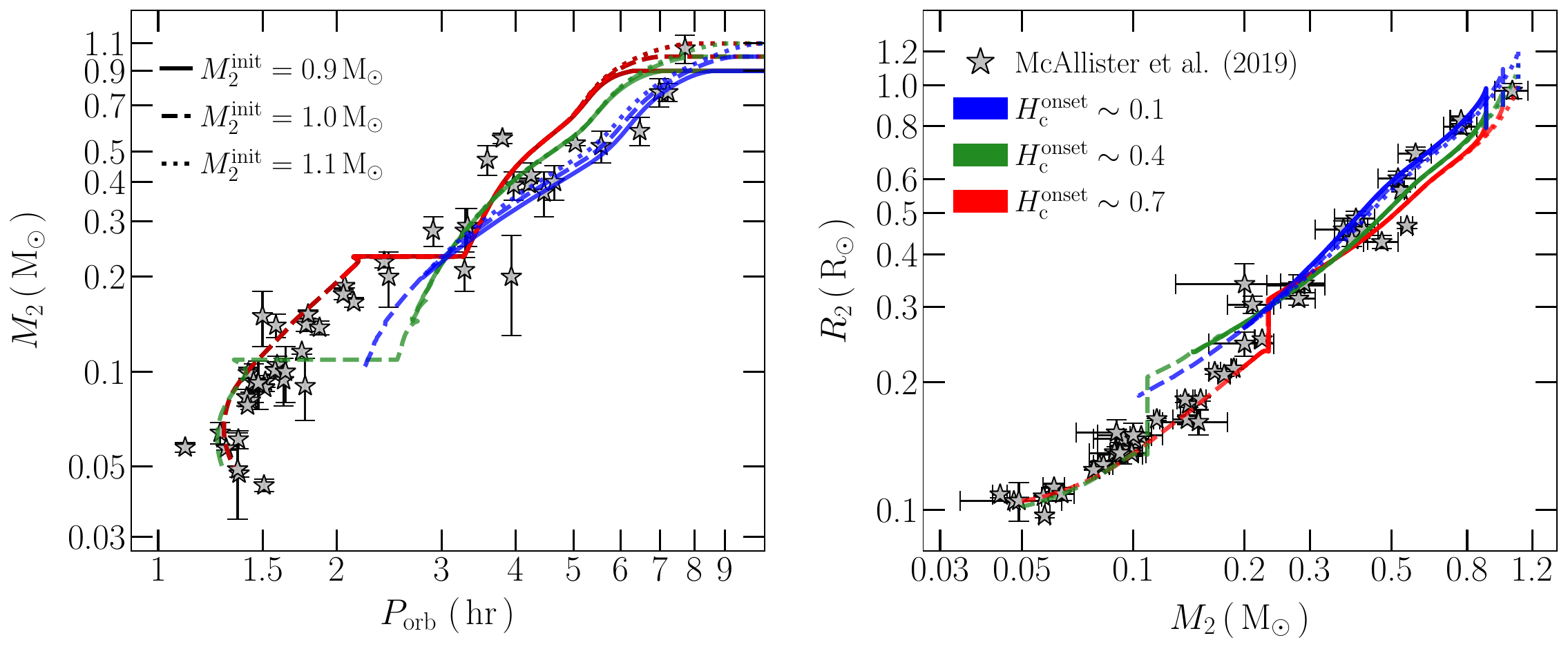}
     \caption{Comparison between predicted (Eq. \ref{eq:MBrecipe}, assuming ${K=\eta=30}$) and observed \citep[gray stars,][]{mcallisteretal19-1} long-period \cv~properties: orbital period against donor star mass (left panel) and donor star mass against donor star radius (right). We assumed a white dwarf mass of $0.83$~\Msun, considered three different initial donor masses ($M_2^{\rm init}$), namely $0.9$ (solid lines), $1.0$ (dashed lines), and $1.1$ (dot-dashed lines), and chose initial orbital periods leading to three different central hydrogen abundances at the onset of mass transfer ($H_{\rm c}^{\rm onset}$), namely ${\sim0.7}$ (green lines), ${\sim0.4}$ (blue lines), and ${\sim0.1}$ (red lines). It is clear from the figure that, regardless of the initial donor mass, the lower the central hydrogen abundance at the onset of mass transfer, the larger the donor star radius. In particular, all long-period systems can be reasonably well explained if the donor star had enough time to evolve that its central hydrogen abundance sufficiently dropped before the onset of mass transfer (i.e., ${H_{\rm c}^{\rm onset}\lesssim0.1-0.2}$).}
     \label{fig:Mdot+R2_solartypedonors}
\end{figure*}

\subsection{Impact of the age of solar-type donors}

We have so far ignored donor stars that are initially more massive than ${\sim0.8}$~\Msun.
However, they seem to be required to explain the observed systems with donor stars more massive than ${\sim0.5-0.6}$~\Msun~since we were unable to explain their radii (as discussed in Sect. \ref{Sect:RVJvsSBD}).
To test this, we ran a set of simulations with initial donor star masses of $0.9$, $1.0$, and $1.1$~\Msun, assuming a white dwarf mass of $0.83$~\Msun.
We adopted the eCAML model, set the metallicity to solar metallicity (${Z=0.02}$), and ${K=\eta=30}$.
The initial orbital periods were chosen so that the central hydrogen abundances of the donor stars at the onset of mass transfer were ${\sim0.1}$, ${\sim0.4}$, and ${\sim0.7}$, with ${\sim0.7}$ corresponding to the assumed initial abundance.

During the evolution of a low-mass main-sequence star, that is, during its central hydrogen burning phase, the central mean molecular weight increases because of the increase in the central helium abundance, while the total number of free particles as well as the pressure in the center decrease.
This causes the core to contract in response, increasing its density, which results in an expansion of the envelope, leading to a larger radius, coupled with an increase in the luminosity (and a slight increase in the effective temperature). 
These features have an important impact on the donor star mass-radius relation during \cv~evolution, as shown in Fig. \ref{fig:Mdot+R2_solartypedonors}.
For a fixed donor mass, the lower the central hydrogen abundance, the larger the radius and in turn the longer the orbital period at which the donor is filling its Roche lobe.
As illustrated in Fig. \ref{fig:Mdot+R2_solartypedonors}, explaining the measured radii of the long-period systems requires donor stars with low central hydrogen abundances (${\lesssim0.1-0.2}$).
Therefore, the SBD magnetic braking model not only explains the systems with orbital periods shorter than ${\sim6}$~hr but also those with longer periods, provided the donor stars are slightly evolved,

\subsection{Dependence on metallicity}

In binary population studies of \cvs~and their progenitors the dependence of \cv~evolution on metallicity is frequently ignored \citep[e.g.,][]{zorotovicetal14-1,goliasch+nelson15-1} while different metallicities can have an impact \citep{stehleetal97-1}. 
Here we varied the metallicity of the donor star to evaluate how different donor star metallicities might affect the predictions of the SBD model. 

\begin{figure*}
    \centering
        \includegraphics[width=1.0
    \linewidth]{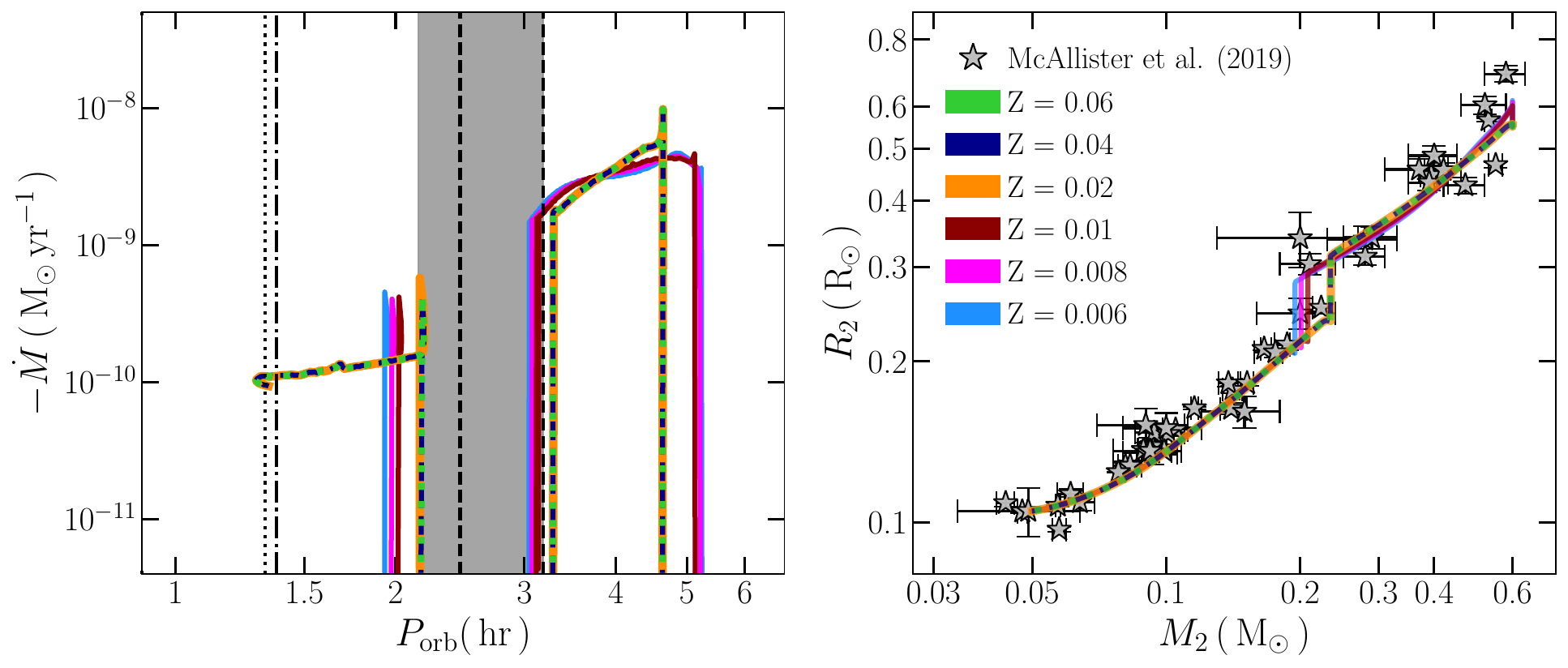}
     \caption{Predicted tracks as a function of the metallicity of the donor star. For low $Z$ ($\lesssim 0.01$) the assumed value of the metallicity changes the predictions while 
     for systems with $Z \gtrsim\,0.02,$ the evolution is essentially independent of the metallicity (the green, blue, and orange lines overlap). We set $K=\eta=30$ and $M_2 = 0.6$\,\msun, while the other initial parameters were set as in Fig. \ref{fig:SBD+RVJ_Jdot+Mdot}.}
     \label{fig:Mdot+R2_Z}
\end{figure*}

In Fig. \ref{fig:Mdot+R2_Z} we show that the evolution of a given system depends on the metallicity of the donor star. 
For low metallicities ($Z\leq0.02$), the lower the metallicity the shorter the orbital periods at which the period gap starts and ends.
For higher metallicities ($Z \geq 0.02$) the period gap starts and ends at identical periods  (longer than in the case of low metallicities) and the tracks are in general nearly indistinguishable. 
The shorter periods at the onset of the period gap for lower metallicities imply lower masses for the donor stars during the detached phase (see the right panel of  Fig. \ref{fig:Mdot+R2_Z})

\section{Discussion}

We have shown that the SBD magnetic braking prescription developed by \citet{bellonietal24-1}, which has been calibrated to explain observations of detached white dwarf plus M-dwarf binaries \citep{schreiberetal10-1} and low-mass main-sequence binaries \citep{elbadryetal22-1}, also convincingly describes the evolution of \cvs~. The boosting and disruption parameters required to produce the period gap and the period minimum are similar to those needed for detached binaries. In what follows we discuss these results in the context of our general understanding of magnetic braking concerning both close binaries and single stars. 
We start by reviewing the performance of the SBD prescription in explaining populations of detached binaries before evaluating possible constraints of the 
critical parameters of the SBD prescription. Finally, we take a look at the empirical convective turnover time used in SBD as well as in several other magnetic braking prescriptions.  

\subsection{Saturated, boosted, and disrupted magnetic braking in detached systems}

The SBD prescription for magnetic braking has been invented to simultaneously explain the observations of 
post common envelope binaries consisting of a white dwarf with an M-dwarf companion \citep{schreiberetal10-1}, which show a drastic increase in the relative number of systems for fully convective companion stars, and main-sequence binaries \citep{elbadryetal22-1}, which provide evidence of a weak dependence of magnetic braking on orbital period for short period binaries, that is, for saturation of magnetic braking. 
\citet{bellonietal24-1} found that the observations can be reasonably well explained by $K=\eta\,\gtrsim\,50$.

In a complementary work, \citet{blombergetal24-1} recently analyzed detached post common envelope binaries with hot subdwarf primaries. Among the observed systems of this type, the dominance of fully convective companions seems to be equally strong as in detached white dwarf plus M-dwarf post common envelope binaries. 
\citet{blombergetal24-1} found that the almost complete absence of hot subdwarf post common envelope binaries with companions that still have a radiative core 
cannot be explained by the \citetalias{rappaportetal83-1} model assuming similar values for $K$ and $\eta$ but that the SBD magnetic braking prescription offers an explanation if $K=\eta\,\gtrsim\,50$. 

Thus, relatively large values of $K$ and $\eta$ for the SBD model are required to explain the companion mass distribution of both types of detached post common envelope binaries. Our individual tracks provide the best agreement with the measured onset of the period gap in \cvs~for slightly smaller values, that is, $K\simeq30$. However, in the following two subsections we provide potential explanations for this difference.

\subsection{Limitations of evolutionary tracks}

 \begin{figure*}
    \centering
        \includegraphics[width=1.0
    \linewidth]{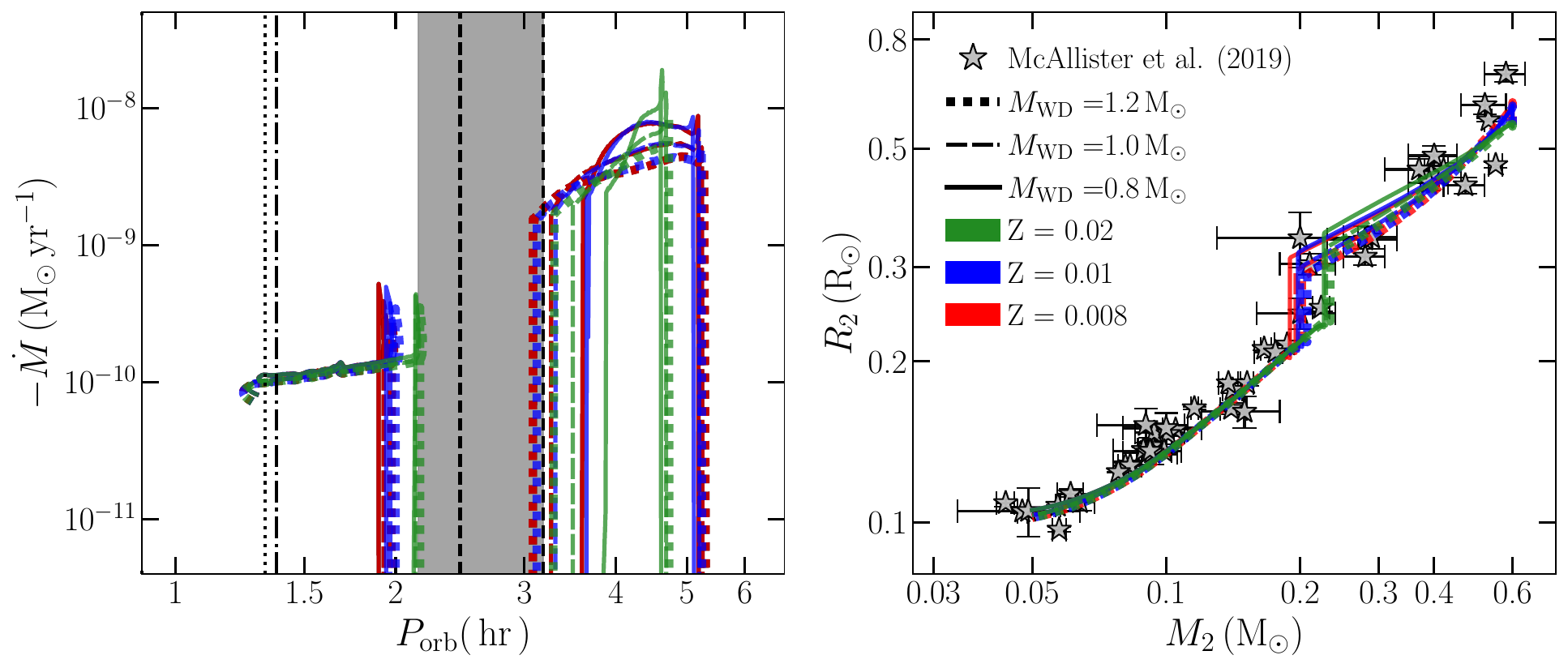}
     \caption{Mass transfer rate evolution and mass-radius relation for different systems using SBD magnetic braking and eCAML for different metallicities and white dwarf masses. The black lines, gray area, and black symbols are as in Fig. \ref{fig:SBD+RVJ_Jdot+Mdot}. The parameters for boosting and disruption were set as $K = \eta = 50,$ and the donor mass and initial period were assumed to be $M_2 = 0.6$ \msun~and $P_{orb} = 1$\,d, respectively. The combination of tracks for different white dwarf masses and metallicities illustrates that both borders of the gap are smeared out. }
     \label{fig:Mdot+R2_MWD+Z}
 \end{figure*}

We show in Sect. \ref{sec:k-eta} that for $K=\eta=30$ both boundaries of the period gap as well as the period minimum are well reproduced assuming a white dwarf mass that corresponds to the average white dwarf mass in \cvs~and for a typical metallicity of $z=0.02$. One thus might naively think that the \cv~orbital period distribution is best reproduced with $K=\eta=30$. However, this represents a simplification because of several reasons. 

First, the observed orbital period gap does not show razor sharp boundaries.  
The boundaries of the gap have been determined visually by \citet{kniggeetal11-1} and \citet{schreiberetal24-1} from different samples of \cvs~with different results for the lower boundary of the gap.
In these samples, the gap itself is not completely void of nonpolar~\cvs. 
This existence of \cvs~in the period gap is expected as systems with fully convective donor stars in the mass range $\sim 0.2-0.35$\,\Msun\,are born inside the gap. 
In addition, depending on the strength of magnetic braking, systems with masses between $0.35-0.4$\,\Msun\, might start mass transfer at periods where systems with more massive donors already detached as a consequence of the disruption of magnetic braking. These systems will experience strong angular momentum loss. As a consequence, they will evolve toward longer periods while the donor is adjusting its radius to the large mass loss rate (an example is the brown track in Fig. \ref{fig:SBD+RVJ_Jdot+Mdot}, middle right panel).  
This means that even for fixed white dwarf mass and metallicity, both boundaries of the gap
will be smeared out. 

Second, of course, white dwarf mass and the metallicity of the donor vary in \cvs,~which affects the predicted boundaries of the period gap. To illustrate this effect we combined evolutionary tracks for different white dwarf masses and donor star metallicities and $K=\eta=50$ in Fig. \ref{fig:Mdot+R2_MWD+Z}. The combination of tracks for different parameters shows that a sharp gap should not be expected and that also for values $K=\eta>30$ one might obtain a good fit. 

The main message here is the following. As a sharp gap is not seen in the observed period distribution and not expected theoretically, we cannot provide very tight constraints on the parameters $K$ and $\eta$ using individual tracks. 
We therefore conservatively estimate that our simulations indicate that $30\,\lappr K=\eta\,\lappr 50,$ which is smaller than but still consistent with the values derived from simulations of populations of detached binaries \citep{bellonietal24-1,blombergetal24-1}. More solid constraints can perhaps be derived by performing population synthesis of \cvs,~which is beyond the scope of the present paper.

\subsection{Relation between convective turnover time and magnetic braking}

A key ingredient of saturated magnetic prescriptions, including SBD, is the dependence of the dynamo number on the convective turnover time and the rotation period \citep{durney+latour78-1}. As the dynamo number of the $\alpha-\Omega$ dynamo depends on the Rossby number, that is, the ratio between rotation period and convective turnover time, activity tracers, such as chromospheric emission or X-ray luminosity, have been related to the Rossby number as well using large samples of low mass stars with known rotation period. 
\citet{noyesetal84-1} determined a relation between chromospheric activity and Rossby number using calculated turnover times \citep{gilman80-1} and a similar approach has been followed more recently by \citet{landinetal23-1}. 
 
In contrast, \citet{wrightetal11} 
used the X-ray luminosities, rotation rates, and colors of $824$ stars to determine the convective turnover time as a function of mass. The resulting semi-empirical relation between stellar mass and convective turnover time has been widely used in theoretical studies including previous works on SBD magnetic braking \citep{bellonietal24-1,blombergetal24-1}. 
We therefore incorporated this turnover time into the MESA calculations presented in this paper.   
The use of an empirical turnover time, however, introduces an observationally determined parameter into our modeling, which holds the potential for an inconsistency.   
Indeed, recent works by \citet{landinetal23-1} and \citet{gossageetal24-1} showed that calculated turnover times deviate from the empirically determined mass dependence \citep{wrightetal11}. 
In addition, the semi-empirical relation seems to be underestimating the masses of late M dwarfs typically by at least a factor of two \citep{jaoetal22-1}. This could especially affect predictions concerning the period gap. 

Finally, and perhaps most importantly, the empirical relation has been made for single main-sequence stars. In \cvs,  
at the onset of mass transfer, the mass loss drives the donor stars out of thermal equilibrium.   
As a consequence \cv~donors are significantly bloated, especially above the orbital period gap when magnetic braking is assumed to be efficient in removing angular momentum. This difference to main-sequence stars causes \cv~donors to become fully convective at a lower mass, that is, at $\sim0.2$\,\Msun~instead of $0.35$\,\Msun~for single main-sequence stars. Using the empirical relation of the turnover time as a function of mass derived from single main-sequence stars for \cv~donors therefore introduces an inconsistency that might impact the values for $K$ and $\eta$ that best fits the period gap and the period minimum.

\subsection{Implications for other close binaries and single stars}

With the development of the SBD magnetic braking prescription for magnetic braking an empirical relation has been found that successfully explains observations of detached white dwarf and hot subdwarf binaries with M-dwarf companions, main-sequence binaries \citep{elbadryetal22-1}, and \cvs. One could therefore argue that the SBD prescription should replace the traditional \citetalias{rappaportetal83-1} model as the standard prescription for close white dwarf binaries. 

However, the consequences of replacing \citetalias{rappaportetal83-1} magnetic braking with the SBD prescription still needs to be tested by performing binary population synthesis. 
Especially because the evolution of \cv~progenitors with donor stars that still have a radiative core will be significantly faster assuming SBD magnetic braking. This means that for the same common envelope efficiency more \cvs~will form. This expected result is a direct consequence of the finding by \citet{elbadryetal22-1}, that magnetic braking also saturates in close main-sequence binaries, that is, it depends little on the orbital period. 

The faster evolution toward the second phase of mass transfer could also imply that assuming SBD magnetic braking leads to predicting the formation of relatively large numbers of AM\,CVn binaries from \cvs~with evolved donors. This is because strong magnetic braking is needed to overcome the fine-tuning problem for AM\,CVn binaries forming from \cvs~with evolved donor stars \citep[for more details see][]{belloni+schreiber23-2}. Similarly, the formation of ultra-compact binaries from low-mass X-ray binaries might be possible. While this needs to be tested by performing dedicated simulations, 
it might well be that SBD magnetic braking could replace the convection and rotation boosted (CARB) 
prescription, which has been successfully used for AM\,CVn binaries \citep{belloni+schreiber23-2} and low-mass X-ray binaries \citep{van+ivanova19-1}. 

Finally, one might also consider applying the SBD prescription to the spin-down of single stars. So far, attempts to unify magnetic braking prescriptions for binaries and single stars have failed dramatically  \citep[e.g.,][]{belloni+schreiber23-1}. A recent example is provided by the attempt to use a magnetic braking prescription derived from field complexity for both binaries and single stars \citep{garraffoetal16-1,garraffoetal18-1,garraffoetal18-2}. While this prescription seemed promising at first glance, more detailed simulations showed that the predicted steep decrease in the magnetic braking efficiency at the fully convective boundary does not predict the existence of a period gap in the period distribution of \cvs~\citep{ortuzaq-garzonetal24-1}. 
We also expect SBD magnetic braking to generate problems when applied to rotation rates of single low mass stars. While detailed simulations applying SBD magnetic braking to single stars still need to be performed, magnetic braking as strong as predicted by SBD for low-mass stars that still have a radiative core, seems to be excluded by observational constraints derived from spin-down rates of single stars \citep[e.g.,][]{newtonetal16-1}. 

\subsection{The impact of white dwarf magnetic fields}

The simulations of \cv~evolution performed here apply to nonmagnetic or weakly magnetic \cvs~only. It has been speculated decades ago that if the accreting white dwarf is strongly magnetic, the efficiency of magnetic braking might be reduced \citep{lietal94-1,webbink+wickramasinghe02-1}. The most recent \cv~populations observationally characterized indeed show that the orbital period distribution of polars, that is, \cvs~with strongly magnetic white dwarf and with the white dwarf rotation being synchronized with the orbital motion, is very different to that of nonpolar \cvs~\citep{bellonietal20-1,schreiberetal24-1}. These findings are in agreement with magnetic braking being reduced due to strong white dwarf magnetic fields. 

Strong white dwarf magnetic fields can complicate the evolution of \cvs~beyond the reduction of magnetic braking. We now know that strong magnetic fields in single white dwarfs appear typically late, that is, when the white dwarfs have cooled for $\sim2$\,Gyr \citep{bagnulo+landstreet21-1}. This late appearance of white dwarf magnetic fields explains the nearly complete absence of strongly magnetic white dwarfs among the detached progenitors of \cvs~\citep{parsonsetal21-1} and might imply that most strong white dwarf magnetic fields in close binaries appear in the semi-detached \cv~phase. 
As shown by \citet{schreiberetal21-1}, this late appearance of strong fields can result in the transfer of spin angular momentum to the orbit, which might cause \cvs~to detach and appear as white dwarf pulsars \citep{marshetal16-1} that then evolve into synchronized detached pre-polars before becoming magnetic \cvs. 

The evolutionary tracks presented in this work ignore entirely the just described impact of strong white dwarf magnetic fields and therefore the presented tracks only apply to nonpolar \cvs. This is acceptable as in particular these systems show the orbital period gap \citep{schreiberetal24-1}. However, future population synthesis of \cvs~should take the emergence of strong white dwarf magnetic fields into account, ideally modeling the origin of strong white dwarf magnetic fields. Currently, two scenarios that appear to be most promising for explaining the origin of white dwarf magnetic fields: a crystallization driven dynamo \citep{isernetal17-1,schreiberetal21-1} or a fossil field created in the convective cores of the progenitor stars \citep{camisassaetal24-1}. 
In any case, incorporating the impact of strong magnetic fields in binary population models of \cvs~will require the use of detailed evolutionary models instead of fast parameterized tools such as the Binary Stellar Evolution (BSE) code
\citep{hurleyetal02-1}.

\section{Conclusions}

The strength and mass dependence of angular momentum loss through magnetic braking is highly uncertain. The predictions of currently available prescriptions vary by several orders of magnitude \citep[e.g.,][]{belloni+schreiber23-1}. 
Typically, different models are used for \cvs~\citep{kniggeetal11-1}, \cvs~with evolved donors \citep{belloni+schreiber23-2}, low-mass X-ray binaries \citep{van+ivanova19-1}, and single low-mass stars \citep{gossageetal23-1}. 

We tested a saturated, boosted (by a factor $K$), and disrupted (at the fully convective boundary using a factor $1/\eta$) magnetic braking prescription. We incorporated this magnetic braking in MESA and calculated \cv~evolution for different stellar masses, metallicities, and boosting and disruption factors, $K$ and $\eta$. The resulting evolutionary tracks illustrate that this magnetic braking prescription reproduces \cv~evolution very well for $K\simeq\eta\simeq30-50$. These values are slightly smaller than but still consistent with the values $K\simeq\eta\gtrsim50$ required for reproducing the fraction of close binaries among detached white dwarf plus M-dwarf binaries as a function of mass \citep{schreiberetal10-1,bellonietal24-1} and the masses of close companions to hot subdwarf primaries \citep{blombergetal24-1}. In addition, given that magnetic braking saturates for periods shorter than $\sim2$\,days, it reproduces the period distribution of main-sequence binaries \citep{elbadryetal22-1,bellonietal24-1}. 

These results make angular momentum loss through SBD magnetic braking a promising candidate for a unified prescription, at least for describing the evolution of close binary stars. To further test this hypothesis, we need to incorporate self-consistently calculated convective turnover times and apply SBD magnetic braking to \cvs~with evolved donors as well as the formation and evolution of low-mass X-ray binaries, and incorporate SBD magnetic braking into \cv~population synthesis. If these additional tests are successful, SBD magnetic braking might provide important constraints that will help us better understand
magnetic dynamos, magnetic fields, and mass loss in low-mass stars. 

\begin{acknowledgements}
MRS and DB are supported by FONDECYT (grant numbers 1221059 and 3220167). DB acknowledges partial support from the São Paulo Research Foundation (FAPESP), Brazil, Process Numbers {\#2024/03736-2} and {\#2025/00817-4}.
\end{acknowledgements}

\bibliographystyle{aa}
\bibliography{bibs}

\end{document}